\documentclass[twocolumn,secnumarabic,amssymb,nobibnotes,aps,prd,showpacs]{revtex4-1}
\usepackage{graphicx}  
\usepackage{dcolumn}   
\usepackage{bm}        
\usepackage{bbm}       
\usepackage{amsmath}
\usepackage{amssymb}   
\usepackage{color}

\newcommand{\plotpoint}[1]{$\,$\protect\includegraphics[height=1ex]{#1}}
\newcommand{\plotpointparan}[1]{(\protect\includegraphics[height=1ex]{#1})}

\begin{document}


\title{Genus dependence of the number of (non-)orientable surface triangulations}
\author{Benedikt Kr\"uger}
\author{Klaus Mecke}
\affiliation{FAU Erlangen-Nuremberg, Institute for Theoretical Physics, Staudtstr. 7, 91058 Erlangen, Germany}
\date{\today}

\begin{abstract}
  Topological triangulations of orientable and non-orientable surfaces with arbitrary genus have important applications in quantum geometry, graph theory and statistical physics.
  However, until now only the asymptotics for 2-spheres are known analytically, and exact counts of triangulations are only available for both small genus and small triangulations.
  We apply the Wang-Landau algorithm to calculate the number $N(m,h)$ of triangulations for several order of magnitudes in system size $m$ and genus $h$. 
  We verify that the limit of the entropy density of triangulations is independent of genus and orientability and are able to determine the next-to-leading and the next-to-next-to-leading order terms. We conjecture for the number of surface triangulations the asymptotic behavior 
  \begin{equation*}
    N(m,h) \rightarrow (170.4 \pm 15.1)^h m^{-2(h - 1)/5} \left( \frac{256}{27} \right)^{m / 2}\;,
  \end{equation*}
  what might guide a mathematicians proof for the exact asymptotics. 
\end{abstract}

\maketitle

\section{Introduction}

Triangulations of manifolds provide a standard method of discretizing surfaces in condensed matter and a possibility to quantize space-time.
They are used in the simplicial quantum gravity models of dynamical triangulations \cite{Ambjorn_1995c}, the causal version thereof \cite{Ambjorn_2005}, as well as in spin-foams \cite{Rovelli_2007}.
Furthermore they are also a fundamental object within the group field theory approach to quantum gravity \cite{Freidel_2005,Oriti_2011,Rivasseau_2012}, which can be seen to relate the previously mentioned approaches.
For the simplicial quantum geometry models it is crucial to know the scaling of the number of triangulations in terms of the system size, because on the one hand the statistical models are only well-defined if there exists an exponential scaling, and on the other hand the scaling constant determines the value of the coupling constant to obtain a phase transition necessary for results independent of the introduced discretization scale \cite{Ambjorn_1995c,Ambjorn_2005}.
Even if triangulations are not seen as a tool for regularization as in simplicial quantum geometry, but as the actual relevant degrees of freedom as in the spin-foam or the group field theory approach, the asymptotics of the number of triangulations in terms of the system size is important for the measure term of the path integral.

For rooted triangulations of the 2-sphere it is well-known that their number scales $\propto \sqrt{256/27}^{m}$, with $m$ being the number of triangles \cite{Tutte_1962}.
A triangulation is rooted by marking some vertex as well as some adjacent edge and face as special in order to break symmetry and to simplify the counting procedure.
For standard triangulations the same result was obtained later by proving that the ratio of triangulations possessing any non-trivial symmetry vanishes for large triangulations \cite{Tutte_1980}.
For other surfaces with different genus or orientability, like the torus or the projective plane, no asymptotic numbers are known, neither for the rooted nor for the default, unrooted case.
Nevertheless, for simplicial quantum gravity triangulations of arbitrary surfaces are important, because the models are not restricted to a certain topology of the underlying manifold.

\begin{figure}
\centerline{
  \includegraphics[width=0.475\columnwidth]{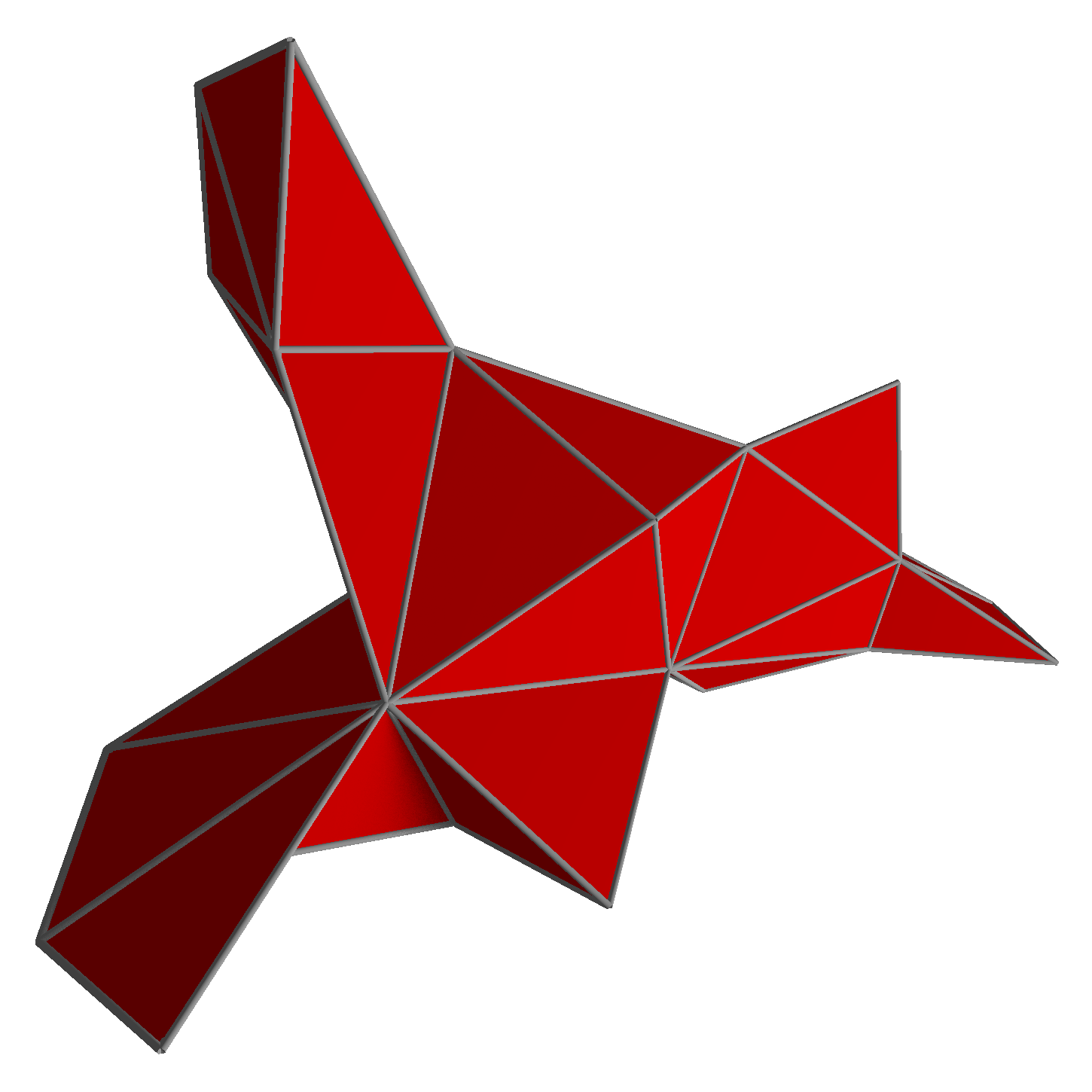}
  \includegraphics[width=0.475\columnwidth]{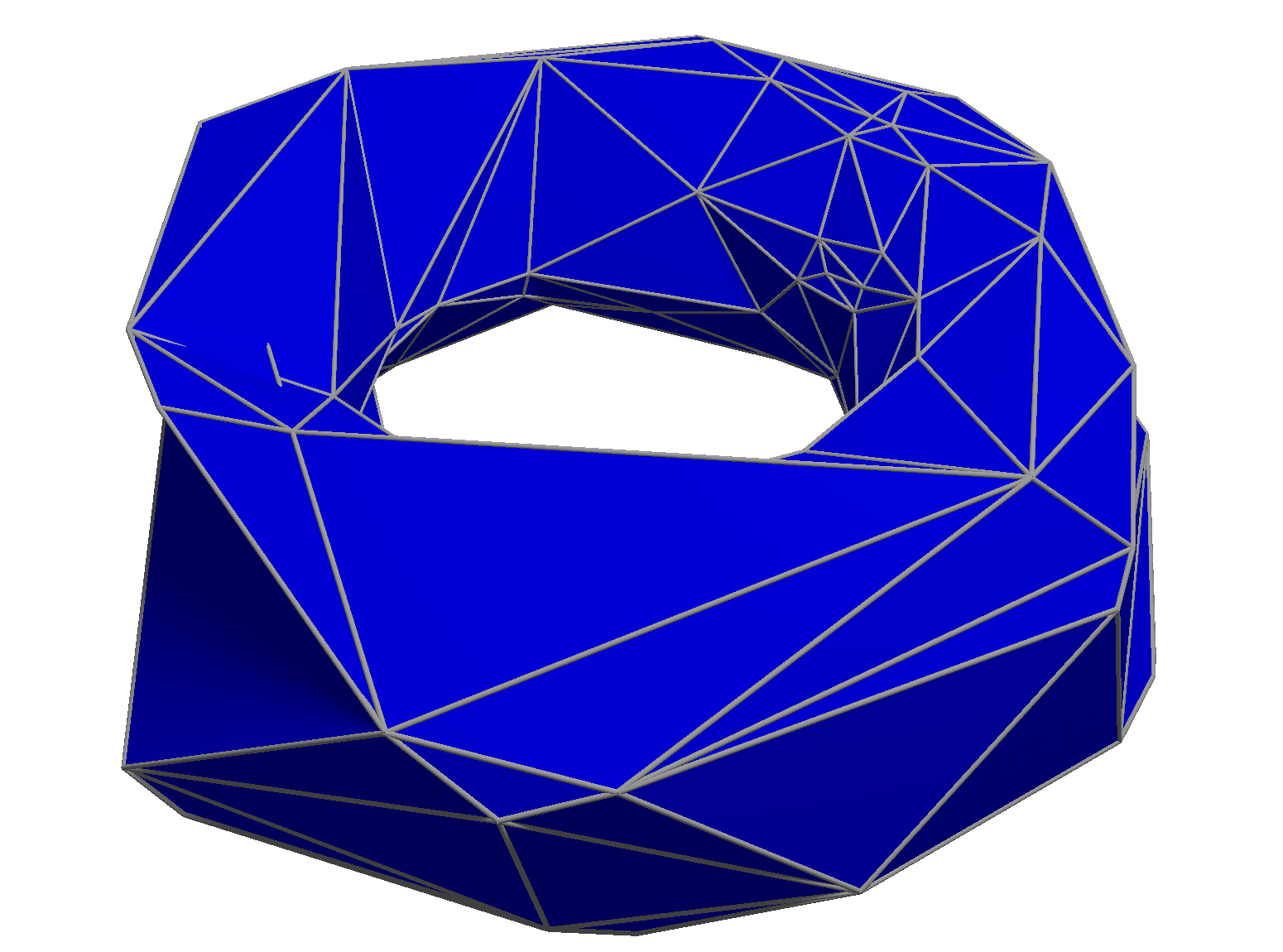}
}
\caption{(Color online) Examples for triangulations of surfaces with low genus. (Left, red) Triangulation of the 2-sphere with $m = 50$ maximal simplices. (Right, blue) Triangulation of the torus (orientable surface with genus $g = 1$ with $m = 200$ maximal simplices.}
\label{fig:triangulation_examples}
\end{figure}

Triangulations of surfaces with non-vanishing genus are also an object of study in other branches of physics:
Since every graph is planar if embedding into a surface with arbitrary high genus, and triangulations are the maximal planar graphs for the respective surfaces (every insertion of an edge would violate the planarity), they are an important tool in graph and network theory \cite{Kownacki_2004,Andrade_2005,Aste_2012}.
Furthermore, critical properties of statistical systems defined on quantum surfaces or triangulated manifolds are sometimes easier to solve than on Euclidean lattices, but can be related to these using the KPZ-formula \cite{Knizhnik_1988,Duplantier_2011,Garban_2013}.

Using lexicographic enumeration it is possible to exactly count triangulations of orientable and non-orientable surfaces for small genus $g \leq 6$ and small number of vertices $v \leq 23$ \cite{Sulanke_2005,Brinkmann_2007,Sulanke_2009}.
For bigger genera or larger triangulations this method does not give results in any reasonable computation time.

In contrast to triangulations, the asymptotic behavior of (triangular) maps on surfaces is far better understood.
A triangular map is a graph drawn on a surface so that each face is a triangle, the main difference to triangulations is that triangular maps allow for digons, multiple edges or loops.
One can show that the asymptotic number $T(k,h)$ (orientable) and $P(k,h)$ (non-orientable) of certain classes of maps on arbitrary surfaces with has the form \cite{Gao_1993}
\begin{equation}\label{eq:asymptotic_maps}
  \left[ \begin{matrix} T(k,h) \\ P(k,h) \end{matrix} \right] = \alpha \left[ \begin{matrix} t_h \\ p_h \end{matrix} \right] (\beta k)^{5(h - 1)/2} \cdot \gamma^k
\end{equation}
where $k$ is the number of edges and $h = g$ (orientable) respectively $h = g / 2$ (non-orientable) is the type of the surface.
The constants $t_h$ and $p_h$ do only depend on $h$ and not on the class of maps that are counted, they were calculated in Ref.\,\cite{Bender_2008} using a recursion relation obtained in Ref.\,\cite{Goulden_2008}.
The numbers $\alpha$, $\beta$ and $\gamma$ depend on the class of surfaces, one finds e.g. $\gamma = 12$ for all maps \cite{Bender_1986} or $\gamma = 2^{2/3}\sqrt{3}$ for triangular maps \cite{Gao_1991}.

In this paper we numerically approximate the number of surfaces triangulation in terms of the genus $g$ and the number of triangles $m$ using the Wang-Landau algorithm \cite{Wang_2001,Wang_2001b} for several orders of magnitude in $g$ and $m$.
A similar version of this method was used in Ref.\,\cite{Knauf_2015} to approximate the number of lattice triangulations.
We are able to extract the long sought asymptotics for the number of triangulations of arbitrary surfaces for the first time and find an exponential growth that coincides with the one found for spheres in Ref.\,\cite{Tutte_1980}. 
Additionally, we determine the sub-exponential corrections similar to Eq. \eqref{eq:asymptotic_maps}, which are a valuable hint for mathematicians proving the exact asymptotics for the number of surface triangulations.
The presented method is not limited to estimating the total number of surface triangulations, but can also be used for estimating the asymptotics of the cardinality for certain subclasses of these triangulations.
A possible application can be estimating the asymptotic number of irreducible triangulations, which are triangulations without contractible edges (see Ref.\,\cite{Barnette_1989} for detailed definition, and Ref.\,\cite{Sulanke_2005} for enumerations of small irreducible triangulations).
Furthermore, our method can also be applied to $k$-equivelar or $k$-covered triangulations, where in the former every vertex has degree $k$ and in the latter there is at least one vertex with degree $k$ (see Refs.\,\cite{Negami_2001,Lutz_2010} for detailed discussion and numbers for few vertices), and to much more different subclasses of triangulations.

\section{Construction of triangulations}

First we present the definition of triangulations of (closed) surfaces $M$ by using the notion of simplicial complexes.
Let $\mathcal I$ be a set and $K \subset 2^\mathcal I$ a set of subsets of $\mathcal I$.
$K$ is an \emph{abstract simplicial complex} if it is complete ($\sigma \in K, \sigma^\prime \subset \sigma \Rightarrow \sigma^\prime \in K$) and closed under the formation of intersections ($\sigma_1, \sigma_2 \in K \Rightarrow \sigma_1 \cup \sigma_2 \in K$).
A \emph{triangulation} $\mathcal T$ of the two-dimensional surface $M$ is an abstract simplicial complex $K$ equipped with an geometric realization (coordinization of every element of $\mathcal I$) that is homeomorph to $M$.
Since the topology of a closed surface is determined only by its Euler characteristic, or equivalently by its genus and orientability \cite{Brahana_1921}, this is also true for the topology of their triangulations.

Triangulations of orientable and non-orientable surfaces with arbitrary genus $g \neq 0$ can be constructed in terms of an arbitrary triangulation of the torus $\mathbbm T$ (orientable surface with $g = 1$) or the projective plane $\mathbbm P$ (non-orientable surface with $g = 1$), using the connected sum $\#$.
Triangulations of the torus and the projective plane can be found e.g. in Ref.\,\cite{Lutz_2011}.
The connected sum of two triangulated surfaces is created by easily be removing a triangle from each triangulation and gluing the boundary together.
A (non-)orientable surface with genus $g$ can then be constructed by taking the g-fold connected sum $\mathbbm T \# \mathbbm T \# \dots \# T$ ($\mathbbm P \# \mathbbm P \# \dots \# P$) of the torus (projective plane).
A triangulation of the 2-sphere, which is the orientable surface with $g = 0$, is given by the boundary of a 3-simplex.

\begin{figure}
\centerline{\includegraphics{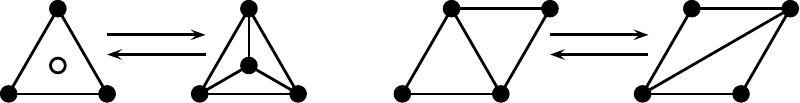}}
\caption{Elementary Pachner-moves in two dimensions. (Left) Insertion and removal of an vertex (Right) Diagonal-edge flip, which is ergodic in the subset of triangulations with the same number of vertices.}
\label{fig:pachner_moves}
\end{figure}

In order to create all possible triangulations of a surface $M$ with given genus and orientability we introduce some elementary steps called \emph{Pachner moves} \cite{Pachner_1986} (which are modification of Alexander moves \cite{Alexander_1930}) that preserve the topology of the underlying manifold $M$ and allow to ergodically create every triangulation of $M$ from every other triangulation of $M$.
In two-dimensions there are three different Pachner moves (see Fig.~\ref{fig:pachner_moves}): 
The first inserts a vertex into a triangle (insertion move), its inverse step removes a three-valent vertex from the triangulation (removal move).
The third step replaces one diagonal of a quadrangle with the other triangle (diagonal-edge move) and is it's own inverse.
In two-dimensions the diagonal-edge moves are ergodic for the subset of triangulations with same number or vertices $v$, if choosing $v$ large enough \cite{Negami_1994,King_2003}.

In order to construct a triangulation of a orientable or non-orientable surface of given genus $g$ and given number of vertices $v$ or number of triangles $m$, we first create a triangulation with the proper genus by taking the connected sum of tori or projected planes as described above.
If the number of triangles in this triangulation is smaller than $m$, we perform insertion moves until the number of vertices equals $m$.
If elsewise the number of triangles is bigger than $m$, we perform removal moves until the triangulation is small enough, and in between diagonal-edge moves if no removal move is possible.
Note that there is a lower bound on the number of vertices or triangles necessary to triangulate a surface with given genus \cite{Heawood_1890,Ringel_1955,Jungerman_1980}, so it is not possible to create arbitrary small triangulations for a given $g$.

\section{Numerical counting algorithm}

We use the Wang-Landau Markov chain Monte Carlo algorithm \cite{Wang_2001, Wang_2001b} to numerically measure the density of states (DOS) $g(m)$, which is up to a normalization factor the number of triangulations with $m$ triangles.
In general a Markov chain Monte Carlo algorithm generates samples $\omega \in \Omega$ from a sample space $\Omega$ according to a given probability distribution $P(\omega)$ by creating a Markov chain of samples with stationary distribution $P(\omega)$.
Therefor the transition probabilities $P(\omega \rightarrow \omega^\prime)$ have to fulfill the detailed balance condition
\begin{equation}\label{eq:detailed_balance}
  \frac{P(\omega_1 \rightarrow \omega_2)}{P(\omega_2 \rightarrow \omega_1)} = \frac{P(\omega_2)}{P(\omega_1)}.
\end{equation}
A famous and often used example is the Metropolis algorithm \cite{Metropolis_1953}, where $P(\omega) \propto \exp[-\beta E(\omega)]$ is basically the Boltzmann factor.
The Wang-Landau algorithm uses the probability distribution
\begin{equation}\label{eq:probability distribution_wl}
  P_{\mathrm{WL}}(\omega) = \frac{1}{g(E(\omega) )},
\end{equation}
so that the probability distribution in terms of the energies is flat, and chooses the transition probabilities
\begin{equation}\label{eq:transition_probabilities_wl}
  P(\omega_1 \rightarrow \omega_2) = \mathrm{min} \left(1, \frac{g(E(\omega_1))}{g(E(\omega_2))} \right).
\end{equation}
For the considered system of triangulations we use as energy of a triangulation $\mathcal T$ its number of triangles $m$, so that every number $m$ of triangles is sampled equally often. 

Naturally the DOS is a prior unknown, as in most physical problems, otherwise the problem of counting the triangulations would already be solved.
So the Wang-Landau algorithm takes an initial estimation of the DOS (in our case a flat distribution $g_{\mathrm{initial}} \propto 1$) and improves it gradually by $g(m) \rightarrow f\cdot g(m)$ whenever a state with $m$ triangles occurs in the Markov chain.
Here $f > 1$ is a modification factor that decreases during the simulation whenever the histogram of visited energies $H(m)$ recorded at this modification factor is approximately flat to ensure the DOS anneals to the correct DOS.
We use $f \rightarrow f^{0.9}$ as decrease for the modification factor, and consider the histogram of visited energies as flat if $\mathrm{min} [H(m)] \geq c \cdot \mathrm{avg}[H(m)]$ with $c = 0.99$ at the beginning of the simulation, relaxing this condition to $c = 0.8$ with decreasing modification factor.
(Note that $H(E)$ is reset after each decrease of the modification factor.)
Our first modification factor is $f = \exp(1)$ and decreases to $f = \exp(10^{-8})$ during the simulation.
Our choice of parameters is way more careful than the original parameters proposed in Refs.\cite{Wang_2001, Wang_2001b}, resulting in a very small statistical error of our results.

\begin{figure}
\centerline{\includegraphics{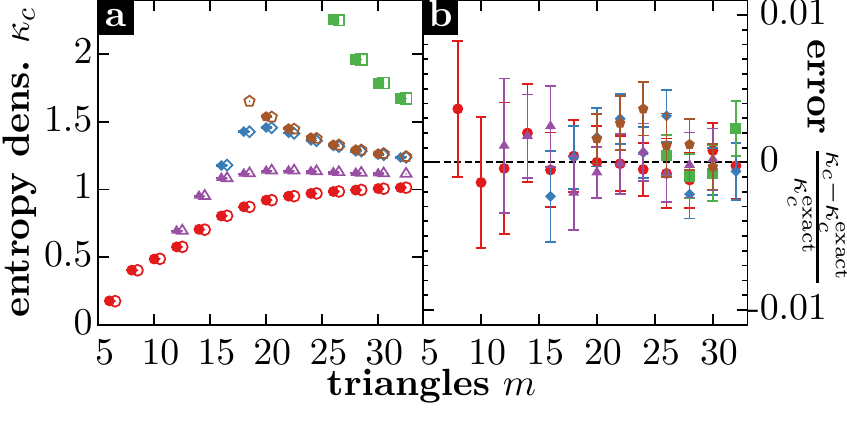}}
\caption{(Color online) Comparison of the exact entropy density from \cite{Sulanke_2009} and our numerical calculations. (a) Entropy density $\kappa_c(m,g)$ in terms of the number of triangles $m$ for orientable surfaces ($g = 0$\plotpoint{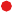}, $g = 1$\plotpoint{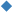}, $g=2$\plotpoint{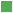}) and non-orientable surfaces ($g = 1$\plotpoint{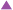}, $g = 2$\plotpoint{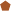}). Our numerical data is plotted with filled symbols, the exact values are plotted with empty symbols and are shifted slightly to the right to resolve these points. (b) Relative error $\kappa_c(m,g) / \kappa_c^{(\mathrm{exact})}(m,g) - 1$ of numerical data with respect to the exact values.}
\label{fig:comparison_exact}
\end{figure}

Instead of counting the number $N_{\mathrm{t}}(m,g)$ of triangulations, we calculate the entropy density $\kappa_c(m,g)$ defined by
\begin{equation}
  \kappa_c(m,g) := m^{-1} \log N_{\mathrm{t}}(m,g).
\end{equation}
By using the entropy density we can use directly the output of the Wang-Landau algorithm, which is the logarithmic DOS, and cancel the normalization factor.
It is also a common quantity discussed in literature \cite{Catterall_1995,Kaibel_2003,Knauf_2015} and corresponds to the value of the (causal) dynamical triangulations' coupling constant for obtaining scale invariance \cite{Ambjorn_1995c,Ambjorn_2005}.
In Fig.~\ref{fig:comparison_exact}a comparison between our calculations and results obtained by lexicographic enumeration for small triangulations \cite{Sulanke_2009} shows excellent agreement and justifies our method.

Due to the diagonal-edge flips being ergodic for large enough surface triangulations it is sufficient to calculate the density of states in an interval $[m - 2, m + 2]$ to obtain the entropy density $\kappa_c$ by
\begin{equation}\label{eq:entropy_density}
  \kappa_c(m) = \frac{1}{8} \cdot \log \frac{g(m + 2)}{g(m - 2)},
\end{equation}
using the assumption that $\kappa_c(m \pm 2) \approx \kappa_c(m)$ valid for large $m$.
Choosing this small interval of computation, the calculation speeds up drastically compared to determining the whole DOS using the Wang-Landau algorithm, because computation time scales with the number of bins.

\begin{figure}
\centerline{\includegraphics{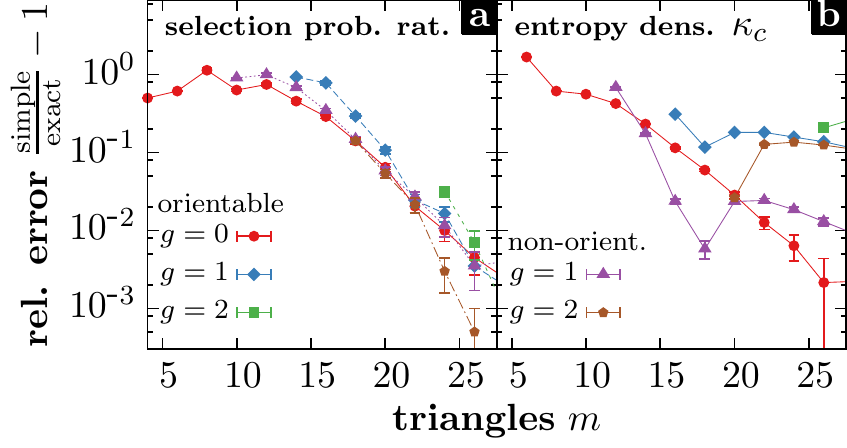}}
\caption{(Color online) Influence of using the simplified selection probability not taking into account isomorphy. (a) Relative error of the selection probability factors for orientable ($g = 0$\plotpoint{cmp_orientable_genus_0.pdf}, $g = 1$\plotpoint{cmp_orientable_genus_1.pdf}, $g=2$\plotpoint{cmp_orientable_genus_2.pdf}) and non-orientable ($g = 1$\plotpoint{cmp_not_orientable_genus_1.pdf}, $g = 2$\plotpoint{cmp_not_orientable_genus_2.pdf}) surfaces in terms of the number of triangles. (b) Relative error of the entropy density.}
\label{fig:selection_probability_error}
\end{figure}

In order to fulfill the detailed balance condition in the Wang-Landau algorithm and to correctly calculate the transition probabilities, one has to compute for each flip the ratio of selection probabilities of the flip and the inverse flip.
Assuming that the triangulation has no special symmetries, the selection probabilities can be determined in terms of the current simplex numbers and their change induced by the flip.
However, there are symmetric triangulations which make it necessary to check whether there are other flips leading to an isomorphic triangulation (these flips are then equivalent), the same for the inverse flip.
These isomorphism checks increase drastically the computation time needed for one step.
But fortunately, as depicted in Fig.\,\ref{fig:selection_probability_error}, the deviations of the exact (with isomorphisms) calculated and the simplified calculations are negligible for triangulations with $m > 30$.
These results are comparable with these of Ref.\,\cite{Richmond_1995} on the level of maps, where it was shown that almost all maps do not posses intrinsic symmetries, which implies in our notion that simplified and exact selection probability match.

\section{Results}

\begin{figure}
\centerline{\includegraphics{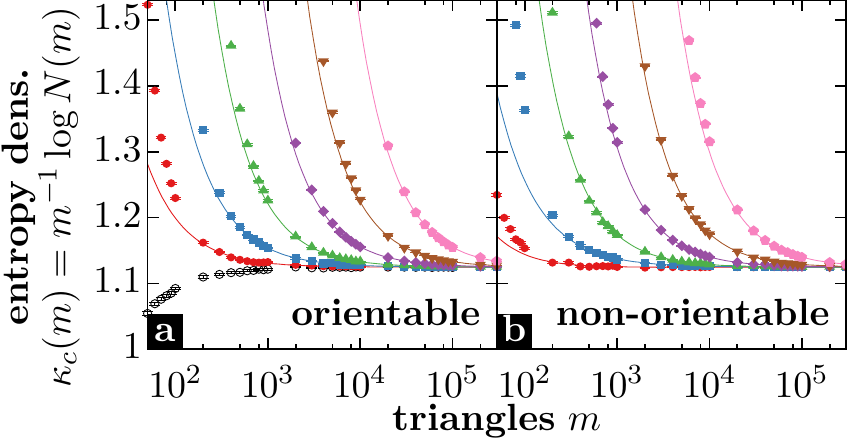}}
\caption{(Color online) Entropy density for triangulations of orientable (a) and non-orientable (b) surfaces in terms of the number of triangles $m$ for genus $g = 0$ \plotpointparan{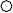} (only orientable), $g = 3$ \plotpointparan{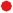}, $g = 10$ \plotpointparan{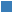}, $g = 30$ \plotpointparan{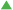}, $g=100$ \plotpointparan{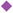}, $g=300$ \plotpointparan{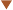} and $g = 1000$ \plotpointparan{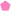}. The lines are fits of Eq. \eqref{eq:entropy_density_scaling} with respect to $a(g)$ and $b(g)$.}
\label{fig:entropy_density_triangles}
\end{figure}

\begin{figure}
\centerline{\includegraphics{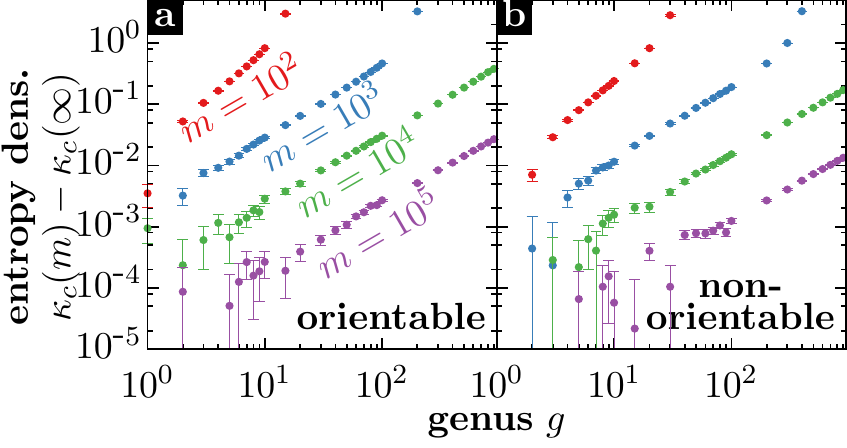}}
\caption{(Color online) Deviation of the entropy density from the limiting value $\kappa_c^{(\infty)}$ for triangulations of orientable (a) and non-orientable (b) surfaces in terms of the genus $g$ for $m = 10^2$ \plotpointparan{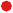}, $m = 10^3$ \plotpointparan{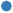}, $m = 10^4$ \plotpointparan{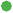}, and $m = 10^5$ \plotpointparan{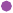} triangles.}
\label{fig:entropy_density_genus}
\end{figure}

We calculated the entropy density \eqref{eq:entropy_density} $\kappa_c(m, g)$ for orientable and non-orientable surface triangulations up to genus $g_{\mathrm{max}} = 1000$ and up to $m_{\mathrm{max}} = 10^7$ triangles using 400 independent Wang-Landau simulations.
In Figs.~\ref{fig:entropy_density_triangles} and \ref{fig:entropy_density_genus} $\kappa_c(m, g)$ is displayed for fixed genus and for fixed number of triangles.

Inspired by the asymptotic enumeration results for triangulations of the 2-sphere and for maps on arbitrary surfaces we assume that the number of triangulations behaves as
\begin{equation}\label{eq:triangulation_number_scaling}
  N(m,g) = \overline{a}(g) \cdot \overline{b}(g)^m \cdot m^{\kappa_c^{\infty}(g)}.
\end{equation}
This implies  for the entropy density \eqref{eq:entropy_density} the relation 
\begin{equation}\label{eq:entropy_density_scaling}
  \kappa_c(m,g) = a(g)\cdot \frac{1}{m} + b(g) \cdot \frac{\log(m)}{m} + \kappa_c^{\infty}(g).
\end{equation}

Considering Fig.~\ref{fig:entropy_density_triangles} we find that the constant term $\kappa_c^{\infty}(g)$ in Eq. \eqref{eq:entropy_density_scaling} does not depend on the genus $g$ and on whether the surfaces is orientable, furthermore we find excellent agreement with the theoretical value of $\log(\sqrt{256/27}) \approx 1.1247$ obtained for triangulations of the sphere \cite{Tutte_1980}.
By inspecting $\kappa_c(m,g = 1)$ one finds that for $g = 1$ the entropy density is approximately constant in terms of $m$, which implies that $b(g) \propto b_1 \cdot (g - 1)$ without any constant term, in agreement to \cite{Tutte_1980} and \cite{Bender_2008}, where $b_1 = 7/2$ for triangulations of the 2-sphere and $b_1 = 5/2$ for triangular maps on surfaces.

\begin{figure}
\centerline{\includegraphics{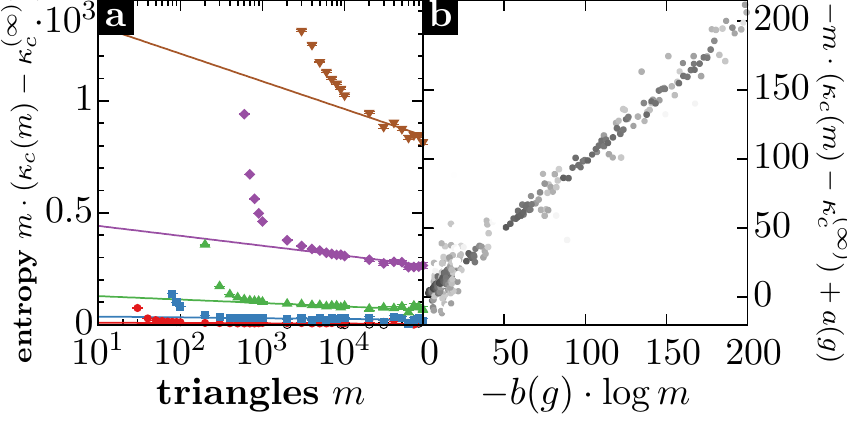}}
\caption{(Color online) Scaling of the entropy density. (a) Plot of the entropy $m \cdot \kappa_c(m)$ in terms of the logarithm of the number $m$ of triangles used to extract the constants $a(g)$ and $b(g)$ by fitting a straight line. The data points are the same as used in Fig.~\ref{fig:entropy_density_triangles}(b) Collapse of the data points, the points are darker for smaller error, only points with $\kappa_c(m) / \kappa_c^{\infty} < 1.05$ are plotted.}
\label{fig:entropy_density_scaling}
\end{figure}

To obtain the constants $a(g)$ and $b(g)$ we rescale the entropy density \eqref{eq:entropy_density_scaling} so that
\begin{equation*}
  m \cdot \left[ \kappa_c(m,g) - c \right] = a(g) + b(g) \log m.
\end{equation*}
For every genus $g$ both constants can then be determined by a linear fit of the rescaled entropy density in terms of $\log m$.
In Fig.~\ref{fig:entropy_density_scaling}a the rescaled entropy density is plotted for orientable surfaces of different genera, one can see an excellent agreement with the proposed linear dependency in terms of $\log m$ for sufficient triangles.
Using the fitted constants all obtained data points can be brought to a collapse as depicted in Fig.~\ref{fig:entropy_density_scaling}b. 

\begin{figure}
\centerline{\includegraphics{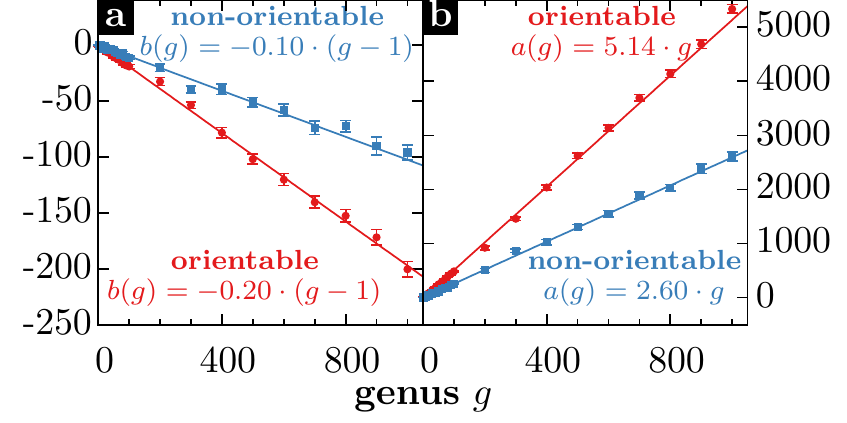}}
\caption{(Color online) Values of the asymptotic constants $a$ and $b$ in terms of the genus $g$ for orientable \plotpointparan{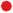} and non-orientable \plotpointparan{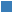} triangulations. The corresponding lines are linear fits.}
\label{fig:asymptotic_behaviour}
\end{figure}

Having fitted $a(g)$ and $b(g)$ for all considered genera of orientable and non-orientable surfaces, one can access numerically the scaling relation of both for triangulations as depicted in Fig.~\ref{fig:asymptotic_behaviour}.
The leading order $b(g) = b_1\cdot (g-1)$ does differ from the results for triangular maps qualitatively, we find $b_1 = -0.197 \pm 0.006$ for orientable and $b_1 = -0.102 \pm 0.004$ for non-orientable triangulations, while for triangular maps the theoretical value $b_1 = 5/2$ was found \cite{Bender_2008}.
We conjecture that $b_1 = -2/5$ for orientable and $b_1 = -1/5$ for non-orientable surface triangulations, since these small integer fractions are in the $1\sigma$ bounds of the numerically obtained values.
The next-to-leading order $a(g) = (5.14 \pm 0.09)\cdot g$ (orientable) respectively $a(g) = (2.60 \pm 0.03)\cdot g$ (non-orientable) has a linear dependency on $g$ for the considered range of genera (implying $\overline{a} \propto \exp(g)$ in Eq. \eqref{eq:triangulation_number_scaling}), no logarithmic correction as proposed in \cite{Bender_2008} for triangular maps is present.
For both $a(g)$ and $b(g)$ one can deduce that the results for orientable and non-orientable triangulations coincide, if one does not consider the genus, but the type of the surface (which is half the genus for non-orientable surfaces).

\section{Conclusion and outlook}
In this paper the number of triangulations of (orientable and non-orientable) surface triangulations with arbitrary genus was calculated using the Wang-Landau Markov chain Monte Carlo algorithm.
Based on our results, we conjecture the following relation for the asymptotic number of surface triangulations
\begin{equation}
  N(m,h) \rightarrow (170.4 \pm 15.1)^h m^{-2(h - 1)/5} \left( \frac{256}{27} \right)^{m / 2}
\end{equation}
in terms of the type $h$ of the surface, which equals the genus $g$ for orientable and half its value for non-orientable triangulations.

These quantitative results for the leading and next-to-leading order terms can be a valuable hint for mathematicians proving the exact asymptotics of the number of surface triangulations.
Additionally, the numerical method presented in this paper can be directly applied to estimate the number and its asymptotics of special types of triangulations, e.g. irreducible, $k$-equivelar or $k$-covered triangulations.

Using the presented method makes it possible to find the scaling behavior of triangulations of higher-dimensional manifolds (either of the total number or of the number of triangulations with certain properties), and to conjecture about fundamental questions that could not be answered until now.
For example, the question whether there are exponentially many or more triangulations of the $d$-sphere (or another underlying manifold) in terms of the number of maximal simplices (facets) \cite{Gromov_2010} can only be answered for certain subclasses of triangulations (e.g., locally constructable triangulations \cite{Durhuus_1995,Benedetti_2011}, geometric triangulations \cite{Adiprasito_2011}, triangulations with a Morse function with a fixed number of critical cells \cite{Benedetti_2012}, or melonic triangulations, which are the dominant contribution to the $1/N$ expansion in group field theory \cite{Gurau_2012}), the answer is there are more than exponentially many in terms of the number of vertices for $d$-spheres \cite{Kalai_1988,Pfeifle_2004,Nevo_2014}.
However, in three and more dimensions the computational effort increases due to the fact that for each data point one has to calculate the DOS $g(m)$ for a much larger interval in the number $m$ of maximal simplices to ensure ergodicity, since there is no similar result as in Refs.\,\cite{Negami_1994,King_2003}.

Our methods and results can also be used for solving simplicial quantum gravity models like (causal) dynamical triangulations on surfaces with arbitrary genus, where the leading order term gives the value of the coupling constant necessary for obtaining scale-independent limits.
Furthermore, the next-to-leading order terms can provide insights into their finite size scaling, which is important since these models are solved mainly using Monte Carlo simulations.

\section*{Acknowledgments}
The authors thank J. F. Knauf for fruitful discussions, as well as B. Benedetti for calling our attention to Ref.\,\cite{Nevo_2014}.
This work is supported by EFI Quantum Geometry and the Elite Network of Bavaria.

\bibliography{literature}

\end{document}